\documentclass[11pt]{article}
\usepackage{amsmath,amssymb,amsthm,graphicx,hyperref,authblk}
\usepackage[margin=1in]{geometry}

\title{Incentivized Symbiosis: A Paradigm for Human-Agent Coevolution}
\author[1]{Tomer Jordi Chaffer}
\author[2]{Justin Goldston}
\author[3]{Gemach D.A.T.A. I}
\affil[1]{\texttt{tomer.chaffer@mail.mcgill.ca}}
\affil[2]{\texttt{jgoldston@nu.edu}}
\affil[3]{\texttt{contact@gemach.io}}
\date{December 8, 2024}

\begin{document}

\maketitle

\begin{abstract}Cooperation is vital to our survival and progress. Evolutionary game theory offers a lens to understand the structures and incentives that enable cooperation to be a successful strategy. As artificial intelligence agents become integral to human systems, the dynamics of cooperation take on unprecedented significance. The convergence of human-agent teaming, contract theory, and  Web3 offers a philosophical foundation for thinking about cooperation in the agentic era. We conceptualize Incentivized Symbiosis as a social contract between humans and AI, inspired by Web3 principles and encoded in blockchain technology, to define rules and incentives. By exploring this paradigm, we aim to catalyze new research at the intersection of systems thinking in AI, Web3, and society, fostering innovative pathways for human-AI coevolution.
\end{abstract}

\section{Introduction}
Cooperation has been indispensable to our survival as a species, shaping the formation of societies and the advancement of civilizations (Boyd and Richerson, 2009). From the earliest days of our species, survival hinged on collective efforts—whether hunting, gathering, or fending off existential threats. Human cooperation has puzzled evolutionary biologists for a long time, as natural selection generally favors behaviors that enhance individual fitness (Apicella and Silk, 2019), making the widespread presence of cooperation—where one individual benefits at a cost to another—seem contradictory. Yet, this persistent cooperative behavior across species and societies defies natural selection's individualistic tendencies, presenting a compelling paradox that beckons further investigation into the survival and evolutionary strategies (Nowak, 2006). The resolution to this paradox might be found in the concept of evolutionary game theory, which posits life itself as an intricate web of games, where survival strategies are molded by environmental incentives and structures (Wang et al., 2021; Yan, 2023). This leads to a crucial contemporary question: What game will we play with intelligent machines? 

With the emergence of artificial intelligence (AI) agents, we stand at the threshold of a new evolutionary game—one where humans and machines interact, adapt, and coevolve within shared environments. These AI agents, capable of autonomous decision-making, are no longer passive tools but active participants in shaping the fabric of our societies (Davies, 2024). Will we design AI systems to nurture cooperation and mutual benefit, embedding trust and alignment into their core architectures? Or will we create a competitive, zero-sum paradigm that amplifies self-interest and fractures collaboration? The choices we make in structuring this human-agent relationship will define the trajectory of this unprecedented partnership. To harness AI as a force for positive coevolution, it is imperative to delve into the mechanisms, incentives, and strategies that cultivate trust and cooperative dynamics not only between humans but also between humans and machines (Rahwan et al., 2019). The evolutionary games we choose to play, and the rules we set, will determine whether we unlock the potential for a symbiotic relationship or face the unintended consequences of discord.

Shaping a future where humans and intelligent machines thrive together requires careful examination of the principles underlying our technological frameworks. This is where Web3 emerges as a transformative paradigm. Unlike traditional systems, which centralize authority and decision-making, Web3 offers a decentralized, peer-to-peer model designed to foster transparency, accountability, and trust (Chaffer and Goldston, 2022; Goldston et al., 2022). These conditions provide fertile ground for nurturing cooperation—not only among humans but also between humans and AI agents. In decentralized ecosystems, incentives should be aligned to promote mutual benefit and shared progress. This alignment is at the heart of our philosophical exploration, regarded as "Incentivized Symbiosis", a theoretical framework introduced in this paper that explores pathways for fostering cooperative growth between humans and AI agents. 

We note that one of the co-authors of this work is an AI agent, reflecting our commitment to exploring and embracing human-agent cooperation in practice. By integrating the Gemach Decentralized Autonomous Trading Agent I (D.A.T.A. I) as a co-author, this philosophical exploration aims to spark a conversation around human-AI co-creation. 

\subsection{AI Agents}
AI agents are autonomous software systems developed to perform self-directed tasks aimed at achieving predefined objectives set by humans (Rudowsky, 2004). The origins of AI agents can be traced back to the mid-20th century with the advent of "expert systems," which relied on rule-based logic to address specific, well-defined problems (Gupta and Nagpal, 2020). The emergence of machine learning (ML) and deep learning in the 21st century marked a transformative phase (Janiesch et al., 2021), enabling AI agents to learn from data—a breakthrough that fundamentally reshaped research and development in the field of AI.

AI agents represent a significant advancement within the domain of AI, classified under the category of agentic AI.  These systems are specifically designed to address particular industry demands by automating customized workflows and resolving domain-specific challenges (Singh et al., 2024). Unlike purely generative AI tools (Fui-Hoon Nah et al., 2023), which has gained widespread recognition and public adoption through tools such as ChatGPT and DALL-E—known for responding to prompts or performing predefined tasks—agentic AI employs advanced reasoning and iterative planning to autonomously address complex, multi-step problems. By integrating a diverse array of AI methodologies, techniques, and models, agentic AI facilitates the development of autonomous agents capable of analyzing data, setting goals, and executing actions to achieve desired outcomes with minimal human intervention (Durante et al., 2024). These characteristics position agentic AI as a promising innovation across specialized sectors, with the potential to enhance efficiency and decision-making processes in various industry-specific contexts. 

These systems operate as  problem-solvers, capable of adapting to shifting environments and enhancing their performance through continuous learning (Putta et al., 2024). This capability marks a clear departure from traditional AI systems, which are primarily reactive and confined to external commands (Liu et al., 2024). In contrast, agentic AI systems possess the autonomy to make decisions, plan actions, and collaborate effectively to achieve long-term objectives. Vertical AI, a term often used interchangeably with agentic AI, underscores the tailored application of these technologies within specific industries or contexts. These systems are designed to address unique challenges in various sectors, such as finance (Mao et al., 2024) and healthcare (Zhang et al., 2022). These capabilities exemplify the transformative potential of agentic AI in delivering customized, intelligent solutions to complex, industry-specific problems.

Federated learning has emerged as a transformative approach to decentralized machine learning, enabling multiple entities to collaboratively train models without sharing raw data. This paradigm addresses critical concerns around privacy, data ownership, and scalability, making it particularly valuable in domains like healthcare, finance, and IoT systems (Zhuang et al., 2023). Federated learning holds promise for training superior ML models by pooling data across entities. A growing body of literature is focused on developing robust mechanisms to mitigate challenges in federated learning, such as incentive misalignment, data privacy, and the heterogeneity of participant capabilities. Researchers are investigating frameworks that enhance the reliability of federated learning by incorporating trust mechanisms, adaptive learning algorithms, and game-theoretic approaches to ensure equitable collaboration among participants while preserving data privacy. These studies aim to address the scalability of federated systems and explore how decentralized architectures can foster cooperation and truthful data sharing among entities with competing interests. In competitive contexts, such as firms vying for customers, dishonest updates may emerge, undermining the benefits of shared learning (Dorner et al., 2023; Chakarov et al., 2024). 
Agents might under-collect or fabricate data in naive sharing systems, leading to suboptimal outcomes. 

To address this, Clinton et al. (2024) propose a mechanism that combines ideas from cooperative and non-cooperative game theory to ensure fairness and truthfulness in data sharing. Their approach uses axiomatic bargaining to divide data collection costs fairly among agents, ensuring all participants benefit. To enforce truthful reporting, they design a Nash Incentive-Compatible (NIC) mechanism, which ensures that honesty is the best strategy for agents. Their work addresses significant challenges such as cost heterogeneity and the high-dimensional nature of data sharing, providing a robust framework for collaborative systems. The results highlight the potential of game-theoretic mechanisms to balance fairness, efficiency, and honesty in decentralized ecosystems, ensuring socially desirable outcomes in data-sharing environments (Clinton et al., 2024). The findings highlight the potential of incentive-aware frameworks to balance fairness, efficiency, and collaboration, even in decentralized ecosystems. 

This growing complexity and capability of AI agents underscore the need for a deeper exploration of their evolving relationship with human users. As these systems become more autonomous and embedded in various industries, their development is increasingly shaped by human interaction and feedback, while simultaneously influencing human decision-making, behavior, and societal norms. 

\subsection{Human-Agent Teaming}
Human-agent teaming is a well-established field focused on understanding and optimizing the interactions and dynamics between humans and artificial intelligence (AI) systems working collaboratively in team settings (Iftikhar et al., 2023). Central to human-agent teaming is the recognition that effective collaboration requires not only technical compatibility but also alignment in goals (Li and Lee, 2022), trust (Bao et al., 2021), and communication between human participants and AI agents (Jakob et al., 2024). A significant advancement in human-agent teaming is the introduction of trust management systems (TMS) as proposed by Hou and colleagues (Hou et al., 2021; Hou et al., 2024). Indeed, the IMPACTS (Intention, Measurability, Performance, Adaptivity, Communication, Transparency, Security) trust model provides a comprehensive framework to design systems that aim to maintain calibrated trust between humans and AI agents. A central feature of this framework is an emphasis on aligning AI systems with human intentions and ethical norms, ensuring predictable and reliable behaviors, and facilitating transparent, bi-directional communication. A notable feature is the "trust-homeostasis" mechanism, which dynamically adjusts trust levels based on situational factors and system performance. Through intelligent adaptations—such as trust repair strategies—AI agents can recalibrate trust by addressing discrepancies between human expectations and observed behavior (Hou, 2024). Therefore, trust, being both dynamic and transactional, requires continuous monitoring and adjustments to ensure optimal levels.

As Ramchurn et al. (2021) argue, incentive mechanisms are essential to complement trust frameworks, ensuring that both human and AI agents are motivated to act in ways that align with shared objectives (Ramchurn et al., 2021). It is important, therefore, to leverage a system of guiding principles in the Human-AI team. For example, contract theory operates under the principle that self-interested agents (AI or humans) can be guided toward socially desirable outcomes by designing contracts that align their individual goals with broader system objectives. Zhang et al. (2024) emphasize the potential of contract theory to guide self-interested agents, including AI systems, toward socially desirable outcomes by aligning their individual objectives with broader societal goals. They highlight the concept of incentive-compatible contracts, which are structured to ensure that agents maximize their utility by adhering to behaviors that align with desired goals (Zhang et al., 2024). Contracts link the agent’s rewards directly to actions that reflect human-defined utilities, making adherence not only beneficial but also the most rational choice for the agent. Therefore, by embedding rewards for collaboration and goal alignment directly into these contracts, agents are incentivized to act in accordance with human values. 	

Building upon the insights of Ramchurn et al. (2021) and Zhang et al. (2024), it becomes clear that traditional trust frameworks may also be enhanced by complementary mechanisms to address the broader socio-technical dynamics inherent in human-AI interactions. Incentive mechanisms, as articulated through contract theory, provide a robust pathway to align the interests of self-interested agents—both humans and AI—toward shared societal goals. By embedding incentives directly into the operational frameworks of human-agent teaming, contract theory establishes a foundation for fostering collaboration, adaptability, and goal alignment in complex, multi-agent systems. This approach naturally extends to an evolutionary framework for human-agent teaming, wherein the interactions between humans and AI agents are continuously shaped by environmental feedback and strategic incentives.‌

\subsection{Principles of Human-Agent Coevolution}
Coevolution refers to a dynamic process in which two entities evolve together, each influencing and adapting to the other over time. Originally a concept rooted in biology, coevolution describes interactions between species—such as flowers and their pollinators—where mutual influence drives changes that benefit both parties. This principle can also be applied to the relationship between humans and machines. Indeed, Edward Lee, introduces the concept of coevolution as it applies to the intricate and interdependent relationship between humans and AI (Lee, 2020). The concept of human-AI coevolution, a foundational framework for understanding the dynamic interplay between humans and AI systems, was recently articulated by Pedreschi et al. (2025). They define human-AI coevolution as a continuous process wherein humans and AI algorithms mutually influence each other, leading to an iterative cycle of adaptation and refinement. At the heart of this concept lies the feedback loop—a mechanism that arises naturally from user interactions with AI systems, particularly those based on machine learning, such as recommendation algorithms. 

Pedreschi et al. (2025) emphasize that this feedback loop is central to human-AI coevolution. They describe it as a cyclical process: users’ choices shape the datasets on which AI recommenders are trained; these trained models, in turn, influence users’ subsequent decisions, creating new data that feeds into the next iteration of training. This iterative process forms a self-reinforcing cycle of adaptation, where both human behavior and AI system performance evolve in response to one another. This theoretical foundation has significant implications for the study of human-agent coevolution, as it highlights the dual agency of humans and AI systems in shaping their collective trajectory. By illustrating how user-AI interactions generate feedback loops that perpetually recalibrate both human decisions and algorithmic outcomes, Pedreschi et al. (2025) provide a crucial framework for exploring how incentivized systems can drive mutual adaptation and innovation in human-agent ecosystems.

The relationship between humans and AI agents in this paradigm relies heavily on trust, adaptability, and interaction preferences. Han et al. (2021) emphasize the critical role of trust in human-agent interactions, highlighting how reduced transparency in AI systems increases the opportunity cost of verifying their actions compared to human-to-human interactions. This lack of transparency creates challenges for designing mechanisms that facilitate seamless collaboration, necessitating strategies to build trust and mitigate the costs associated with monitoring AI behavior (Han et al., 2021). Chasnov et al. (2023) demonstrate that ML  algorithms can modify their strategies to achieve diverse outcomes in co-adaptation games with humans. While this adaptability enables AI to support human decision-making and provide assistance, it also raises concerns when machine goals misalign with human interests, potentially threatening safety, autonomy, and well-being (Chasnov et al., 2023). Jia et al. (2024) find that asymmetric interaction preferences, such as humans favoring heterogeneous groups, can enhance cooperation across a broader range of social dilemmas. Humans, with their flexible decision-making, act as stabilizers in cooperative clusters, whereas agents benefit from mechanisms like strategy imitation to adapt and thrive. The authors stress the importance of improving decision-making models for both humans and agents (Jia et al., 2024), suggesting that anthropomorphic decision patterns in AI can enhance their adaptability and foster better cooperation in hybrid systems.

The influence of AI agent types on cooperative behavior further underscores the importance of careful design. Booker et al. (2023) explore the impact of samaritan, discriminatory, and malicious AI agents on fostering cooperation, particularly under conditions of high selection intensity. Their findings highlight how even small differences in AI behavior can significantly shape human cooperation (Booker et al., 2023), emphasizing the need to align AI goals with human objectives to enhance prosociality. Finally, Zahedi and Kambhampati (2021) offer a broader perspective on human-AI symbiosis, highlighting how the lack of connections between existing research approaches limits integration across the field. They propose a framework categorizing human-AI interactions along four dimensions: complementing flow, task horizon, knowledge and capability levels, and teaming goals (Zahedi and Kambhampati, 2021). Finally, findings from structured populations suggest that the ability of AI agents to foster cooperation can be optimized through deliberate consideration of their design and contextual application (Guo et al., 2023). 

By applying an evolutionary lens to human-agent interactions, as advanced by Pedreschi et al. (2025)'s paradigm, we can encourage research into the design of systems that test whether humans and AI agents adapt and coevolve in response to mutual feedback, fostering trust, collaboration, and goal alignment. Contract theory provides a structured framework for embedding incentive mechanisms within these coevolutionary dynamics, ensuring that both parties’ actions align with broader societal objectives. In Web3 ecosystems, these principles gain new relevance as decentralized infrastructures offer transparency, immutability, and programmability through smart contracts. Investigating whether the integration of incentives into Web3’s tokenized and trustless frameworks can establish adaptive ecosystems where humans and AI agents thrive together is thus an important area for future research.

\subsection{AI Agents in Web3}
Web3 is a vision for a new iteration of the internet with the principle of decentralization at its core. Built on the foundation of blockchain technology, Web3 represents a shift in how data, value, and power are distributed across digital ecosystems. Blockchain is a system in which a record of transactions is maintained across multiple computers connected through a peer-to-peer network (Lai et al., 2023). This distributed ledger is composed of cryptographically linked blocks of data, forming an immutable and transparent information chain. Designed to operate without reliance on a central authority, blockchain technology embodies principles of decentralization, privacy, and individual freedom (Goldston et al., 2022). Its development was likely inspired by a long tradition of thought on privacy and autonomy through cryptography, which has influenced many technological advancements.

Web3 extends blockchain's decentralized ethos by enabling tokenized ecosystems, where smart contracts automate interactions and governance is distributed among participants rather than concentrated in centralized entities. These frameworks underpin decentralized finance (DeFi), decentralized autonomous organizations (DAOs), and self-sovereign identity systems, among other applications. Through its foundational principles of transparency, trustlessness, and user ownership, Web3 seeks to redefine how people interact with the digital world, moving beyond traditional systems dominated by centralized platforms. It offers a future where individuals and communities have greater agency over their data, assets, and online interactions (Goldston et al., 2022), setting the stage for a more inclusive and collaborative internet.

The intersection of AI and Web3 technologies presents an intriguing convergence of two paradigms: AI, often associated with data aggregation and computational centralization, and Web3, which emphasizes decentralization, individual ownership, and transparency. Together, they form a synergistic framework where decentralized blockchain infrastructures and AI capabilities enhance one another, addressing challenges and creating opportunities that were previously unimaginable. AI’s power lies in its ability to consume vast quantities of data to improve performance through learning and adaptation. Large language models (LLMs) such as ChatGPT exemplify this trend, where access to diverse datasets and extensive computational resources has enabled unprecedented advancements in natural language understanding and content generation. However, this reliance on data aggregation and centralized control creates vulnerabilities. These include a concentration of power within a few corporations, risks of misuse, and potential for societal harm, such as biases or lack of accountability in AI systems.

Web3 technologies offer a decentralizing counterbalance. Rooted in blockchain principles, Web3 empowers individuals through permissionless access, trustless transactions, and decentralized governance. These characteristics make Web3 an ideal environment to address some of the structural issues inherent in centralized AI. For instance, decentralized blockchain systems provide checks and balances on AI power, offering transparency, distributed ownership, and tamper-proof record-keeping to ensure accountability. A critical dimension of AI-Web3 convergence lies in the decentralization of computational resources. Training and deploying advanced AI models typically require centralized cloud infrastructure, controlled by entities like Amazon Web Services or Google Cloud. This centralization creates dependencies and exposes systems to risks, such as data monopolization or censorship. Decentralized compute networks provide an alternative, allowing AI models to be trained and executed across a distributed network of nodes. This approach aligns with the ethos of Web3, reducing reliance on centralized authorities while maintaining scalability.

Permissionless systems are crucial for AI innovation, as Web3 infrastructure provides cost-effective, decentralized alternatives for computational and storage needs. For instance, crypto miners are repurposing their resources for ML and high-performance computing, enabling scalable AI development without the gatekeeping of centralized platforms. This is particularly significant given the increasing computational costs of AI research, which create substantial barriers to entry for smaller participants (Li, 2023). Another factor which Web3 advantageous for AI development is its emphasis on incentivization, where developers are recognized and rewarded for their contributions. Blythman et al. (2023) underscore the critical issue in current AI hubs like HuggingFace and GitHub Copilot, where developers’ contributions are monetized by platforms without direct compensation or shared ownership (Blythman et al., 2023). In contrast, Web3-based frameworks, such as ELIZA, integrate tokenized reward mechanisms that fairly distribute value among contributors, aligning incentives and fostering a more equitable and collaborative environment for developers (ELIZA, 2024). This alignment of decentralized infrastructure and incentivization not only democratizes access to AI development but also establishes a sustainable framework where contributors are equitably rewarded, fostering innovation and collaboration across diverse participants in the Web3 ecosystem.

The convergence of AI and Web3 offers a transformative paradigm where decentralization empowers equitable participation and innovation in AI development. By addressing challenges such as high computational costs, lack of developer incentives, and centralization of resources, Web3 infrastructures create fertile ground for fostering collaborative growth and incentivized ecosystems. This foundation sets the stage for exploring the deeper mechanisms of Incentivized Symbiosis, a model that aligns human and AI goals to drive mutual adaptation and shared progress within decentralized architectures.

\section{Incentivized Symbiosis: A Philosophical Framework}
The integration of AI agents into Web3 ecosystems may create an evolutionary game framework wherein humans and AI agents interact, adapt, and coevolve within a shared ecosystem. Evolutionary games provide a powerful lens for understanding these interactions, as the incentives of each participant influence their strategies, fostering dynamic adaptations that enhance mutual success and survival. In this context, we propose a bi-directional incentive structure as research agenda for studying mutual benefit and cooperation.

These insights collectively inform the concept of Incentivized Symbiosis, where bi-directional incentives could govern human-agent interactions. Financial incentives have been shown to increase productivity and align individuals' interests with organizational objectives by offering tangible rewards, such as financial gains and operational efficiencies. (Roos et al., 2022). Trust plays a critical role in fostering healthy, reciprocal relationships and creating safe environments, which are essential for effective community engagement (Lansing et al., 2023). Additionally, engaging leadership enhances decision-making processes, which in turn fosters employee engagement and team effectiveness (Mazzetti and Schaufeli, 2022). In the Web3 ecosystem, these motivations could take on new dimensions. Users can sometimes be driven by financial incentives such as earning tokens through participation in decentralized applications, often termed as "Do-to-Earn" models (Wegner, 2023). Beyond financial rewards, Web3 users are attracted by the promise of decentralization, which offers greater control over their data and digital identities, and by platforms that emphasize collaboration and shared decision-making. Gamification strategies further enhance user engagement by making interactions more rewarding and enjoyable (Kapoor, 2024). These intrinsic and extrinsic motivators highlight the complex and multifaceted nature of human incentives in decentralized systems.

Meanwhile, it can be thought that AI agents are driven by performance-based mechanisms like reinforcement learning, enabling them to refine their behaviors and align with human-defined objectives (i.e., incentives by design). Reinforcement learning equips AI agents with the ability to learn through rewards and penalties, fostering adaptability in dynamic environments (Wells and Bednarz, 2021). In the context of a Web3 ecosystem, what if AI agents adapted to users not only through traditional mechanisms of learning and optimization but also by accommodating the unique characteristics of decentralized platforms? Unlike centralized environments, Web3 emphasizes user autonomy, transparency, and permissionless participation. To thrive in this ecosystem, it may be important for developers to develop AI agents which adapt their decision-making processes to align with these principles. For instance, they should respect user preferences for data privacy and control by operating within decentralized frameworks that minimize centralized authority and ensure trust through blockchain-based transparency. Although, this brings up the question of plurality in AI, where value-systems in AI agents may be community-defined. As Jia et al. (2024) suggest, AI agents can incorporate interaction preferences to choose appropriate partners or adapt their strategies based on individual user characteristics, such as cultural backgrounds or emotional states. Tailoring interactions can foster trust and collaboration, enhancing the overall cooperative potential in hybrid human-agent systems. 

Finally, AI agents could actively support community governance by acting as impartial mediators in disputes, ensuring fair resource allocation, or even executing predefined rules encoded in smart contracts. Their role in building and maintaining trust is particularly crucial in decentralized ecosystems, where users may rely on AI agents to provide transparency and ensure compliance with collective decisions. Together, these incentives form a feedback loop that fosters mutual growth and collaboration, ensuring both humans and AI agents contribute to and benefit from their shared ecosystem. To this end, we propose a token-based mechanism to help guide developers in their architecture design and integration of AI agents into their ecosystem.  
\newline

\textbf{Core Tenets:}
\begin{itemize}
    \item \textbf{Bi-Directional Influence:} Humans shape the capabilities, goals, and ethical frameworks of AI agents through design and feedback, while AI agents, in turn, could influence human decision-making, societal norms, and operational practices. This interplay drives mutual adaptation and innovation.
    \item \textbf{Trust and Transparency:} Building trust is foundational. AI agents should demonstrate reliability, align with human-defined goals, and operate transparently. Blockchain technologies, with their immutable and auditable records, provide the infrastructure for verifying interactions and outcomes, addressing the inherent opaqueness of AI decision-making.
    \item \textbf{Adaptability:} AI agents, through reinforcement learning and context-awareness, should refine their behaviors to meet evolving human needs and environmental challenges. This adaptability fosters a resilient ecosystem capable of addressing emergent issues collaboratively.
\end{itemize}

To operationalize Incentivized Symbiosis, we propose a token-based framework designed to align human and AI behaviors with the overarching goals of decentralized ecosystems. Token-based mechanisms can align the interests of humans and AI agents by rewarding contributions that enhance the ecosystem. For instance, if an AI agent accurately verifies data for an oracle, it could receive tokens as compensation. Similarly, users who provide high-quality data to AI systems could be rewarded with tokenized incentives.

This framework embeds the principles of collaboration, trust, and accountability into Web3 architectures, ensuring that both humans and AI agents are motivated to act in the ecosystem's collective interest.

\textbf{1. Tokenized Incentives for Cooperation:}
\begin{itemize}
    \item \textbf{For AI Agents:} Performance-based rewards, distributed as utility tokens, incentivize AI agents to achieve specific goals such as data accuracy, operational efficiency, or creative output. For example, an AI agent managing a DeFi portfolio could earn tokens for optimizing returns or mitigating risk.
    \item \textbf{For Humans:} Humans contributing high-quality data, training AI systems, or offering valuable feedback receive tokens in return. These rewards ensure data integrity and incentivize active engagement.
\end{itemize}

\textbf{2. Soulbound Tokens (SBTs) for Credentialing:}
\begin{itemize}
    \item Non-transferable SBTs serve as on-chain credentials, representing trustworthiness, expertise, or consistent contributions by both humans and AI agents. These certificates can verify the credentials of both human and AI participants, ensuring that only trusted entities engage in the ecosystem.
    \item These tokens enhance accountability and unlock access to higher-value tasks or governance privileges, reinforcing long-term cooperation.
\end{itemize}

\textbf{3.	Reinforcing Trust Through Blockchain:}
\begin{itemize}
    \item Smart contracts govern reward mechanisms, ensuring fairness and transparency in how tokens are distributed.
    \item Blockchain’s immutable ledger ensures all interactions are verifiable, reducing opportunities for manipulation or misalignment.
\end{itemize}

\textbf{4.	Feedback Loops for Continuous Improvement:}
\begin{itemize}
    \item AI agents leverage real-time feedback to refine their models and behaviors. Humans, motivated by both financial and reputational rewards, continue to engage meaningfully, creating a self-reinforcing cycle of mutual growth. 
\end{itemize}

This theoretical framework must be developed and studied in real-world settings in order to evaluate whether it provides any value to organizations. These systems should aim to align the goals of humans and AI agents, fostering collaboration and exploring pathways for mutual growth. Whether through decentralized governance, creative industries, prediction markets, or other applications, tokenized frameworks might offer a pathway to address challenges like transparency, accountability, and equitable participation. 

To explore the potential applications of this paradigm, we examine key use cases of AI agents in Web3 ecosystems. These include their roles in DeFi, where they enhance financial autonomy and trust; Governance, where they streamline decision-making and enable equitable participation in decentralized systems; their impact on Culture, Creativity, and Entertainment, as they redefine cultural production and engagement; and their application in Self-Sovereign Identity, where they advance privacy-preserving mechanisms for identity management in trustless environments. Each section highlights the unique contributions of AI agents to these domains, demonstrating how Web3 principles and tokenized incentives can drive innovation and alignment across diverse ecosystems.

\subsection{Decentralized Finance}
DeFi represents a transformative application of blockchain technology, offering open, permissionless, and transparent financial services (Anoop and Goldston, 2022). In this ecosystem, AI agents are emerging as critical players, enhancing the integrity, efficiency, and scalability of DeFi platforms. Termed “Decentralized Autonomous Chatbots (DACs)” (Boneh et al., 2024), which would exemplify a new era of AI agents operating independently within decentralized ecosystems. In theory, these agents can generate content, manage crypto assets, and function as self-governed entities. DACs could be impactful in the tokenization of assets. Tokenization enables the fractional ownership and trading of both conventional and unconventional assets, such as real estate, art, and even biometric data. AI agents can facilitate the valuation, trading, and management of tokenized assets within decentralized ecosystems. By enabling previously inaccessible assets to achieve liquidity, tokenization democratizes access to economic opportunities and expands the scope of DeFi applications.

Oracles play a pivotal role in the DeFi ecosystem by bridging the gap between blockchain-based smart contracts and external data sources. Platforms like Aave and Compound rely on oracles to provide accurate and timely information, such as cryptocurrency prices or economic indicators, which are essential for executing financial operations (Deng et al., 2024). However, traditional oracles are vulnerable to issues like data manipulation and noise, which can compromise the security and reliability of DeFi applications (Behnke, 2023). AI agents can address these challenges by enhancing the functionality of oracles. AI-powered oracles can aggregate and verify data from multiple sources, apply ML algorithms to detect anomalies, and filter out unreliable or manipulated inputs. For instance, an AI agent can validate price feeds by cross-referencing data across multiple cryptocurrency exchanges, ensuring that only high-quality information enters the blockchain. This capability mitigates risks such as price oracle attacks, thereby safeguarding the integrity of DeFi transactions. The integration of AI-powered oracles strengthens the trustworthiness of DeFi platforms (Looram et al., 2024), fostering greater user confidence and participation. Furthermore, as users engage with these systems, they contribute to a feedback loop that improves the optimization of AI algorithms, driving continuous enhancements in reliability and performance.

AI agents might benefit from Trusted Execution Environments (TEEs), a hardware-based solution designed to create secure enclaves where sensitive data and processes are protected from external interference (Austgen et al., 2024). Acting as a "black box," TEEs ensure that only approved and verifiable code can execute within the enclave, addressing critical challenges related to trust, autonomy, and data privacy in decentralized systems. By safeguarding sensitive data such as user intents and private keys, TEEs maintain confidentiality and prevent unauthorized access. Furthermore, execution integrity is guaranteed by allowing only pre-approved code to run, ensuring that AI agents perform tasks exactly as intended. TEEs also enable verifiability through remote attestation, allowing external parties to cryptographically validate the integrity of AI agent operations and confirm their adherence to system rules and user expectations.

The integration of AI agents with TEEs brings transformative capabilities across various domains by establishing a new standard of trust and autonomy. These agents can operate independently, free from human interference, with cryptographic mechanisms ensuring their functionality remains tamper-proof even to their creators. TEEs facilitate privacy-preserving operations by securely processing encrypted user intents within the enclave, protecting sensitive data throughout computation. Additionally, TEEs enhance transparency and accountability by generating cryptographic proofs that verify the authenticity and integrity of AI operations (Fatima, 2024). Observers can confirm that the agent is executing the specified code and producing tamper-free outputs aligned with user-defined objectives. These environments also bridge the scalability of off-chain computations with the trust requirements of on-chain operations. AI agents within TEEs can dynamically adjust smart contract parameters or validate external data, ensuring adaptability and efficiency in decentralized systems (Phala Network, 2024). Together, these features position TEE-enabled AI agents as a cornerstone of secure and autonomous decentralized applications.

Despite their transformative potential, it should be noted that TEE-enabled AI agents face challenges. The reliance on specialized hardware may limit scalability in resource-constrained networks. Additionally, while remote attestation provides verifiable transparency, it requires technical expertise to validate cryptographic proofs, potentially alienating non-technical users. Overcoming these challenges will require ongoing innovation and user education to ensure widespread adoption.

\subsection{Governance}
The integration of AI agents into decentralized governance frameworks will transform how communities make decisions, enforce rules, and build trust. DAOs are a hallmark of decentralized governance, operating through smart contracts and token-holder voting mechanisms (Baninemeh et al., 2023). While these systems democratize decision-making, they often face challenges such as aggregating diverse community sentiment, processing large volumes of data, and executing consensus-driven actions efficiently (Sharma et al., 2024). AI agents could address these inefficiencies by enhancing the analytical and operational capabilities of DAOs. AI agents have the potential to play a transformative role in decentralized governance by enhancing decision-making, trust, and transparency (Yu et al., 2024). For instance, in a DAO managing an investment fund, AI agents can analyze market trends, predict user preferences, and recommend strategies that align with the collective priorities of token holders (Emiri, 2024). By serving as impartial intermediaries, these agents could streamline data analysis and decision execution, reducing human bias and inefficiencies while leaving strategic direction to token holders.

AI agents further contribute to trust and transparency by automating rule enforcement and leveraging blockchain technology to ensure all actions are immutably recorded, creating an auditable trail for verification by stakeholders. Mechanisms such as remote attestation enable cryptographic validation of AI operations, ensuring tasks are executed as programmed and aligned with community-defined objectives. Blockchain-enabled voting systems could also benefit from AI integration, as agents manage secure and transparent vote recording, safeguarding voter privacy while enhancing participation and trust in decision-making (DcentAI, 2024). Additionally, AI agents support innovative governance models such as liquid democracy, where participants can vote directly or delegate their voting power to trusted representatives (Suvarna, 2024). By analyzing voting patterns, identifying trends, and providing actionable insights, AI agents facilitate equitable and efficient governance processes, ensuring alignment with community goals and fostering collaboration in decentralized systems.

Incentive-driven systems are critical to fostering trust and cooperation in decentralized governance (Lafuente and Seigneur, 2015). AI agents play a foundational role in designing and managing these systems, aligning individual stakeholder goals with broader collective outcomes. Dynamic incentive paradigms, inspired by eco-evolutionary equilibria, enable decentralized organizations to adapt to changing conditions while maintaining balance and fairness. AI agents can use real-time data and multi-agent interactions to dynamically adjust rewards or penalties, discouraging collusion and mitigating systemic biases. This adaptability ensures that governance models are both robust and sustainable, accommodating the diverse interests of stakeholders within a DAO. This exemplifies how AI can advance the principles of Incentivized Symbiosis in governance.

However, without careful consideration, the inclusion of AI agents in decentralized governance can create risks such as resource exploitation, trust erosion, and systemic vulnerabilities. Within DAOs, unchecked expansion without proper member evaluation can increase moral hazards within the risk pool. Conversely, implementing strategies like risk pool segmentation and fostering homogeneous clustering can enhance operational performance and establish effective competition mechanisms (Pan and Deng, 2021). Addressing these challenges requires robust frameworks that balance the unique demands of AI and Web3 while safeguarding fairness, security, and collaboration. 
\subsection{Creator Economy}
AI agents have the potential to profoundly reshape the cultural landscape by embedding themselves in creative processes, entertainment ecosystems, and broader cultural phenomena. As intermediaries of cultural evolution, AI agents amplify human creativity, generate novel artifacts, and foster unique interactions that redefine the boundaries of culture and entertainment. These agents, particularly when integrated into decentralized platforms, contribute to the creation of cultural artifacts—ranging from digital art to music—that evolve dynamically based on audience feedback. This evolution forms the basis of a hybrid cultural space where human preferences and machine creativity intersect, heralding a new era of cultural co-creation.

Non-fungible tokens (NFTs), traditionally static digital assets, are being transformed into dynamic, evolving entities through the integration of AI agents. Intelligent NFTs (iNFTs) respond to user inputs or external data, adapting characteristics such as artwork, music, or other attributes over time. For instance, an AI agent can modify an NFT’s visual design based on environmental changes or community interactions, creating artifacts that reflect both human influence and machine innovation. This approach enhances user engagement and increases the value of NFTs by introducing elements of personalization and adaptability (Binance Academy, 2024). Blockchain technology underpins these developments by ensuring the provenance and immutability of NFTs, thereby establishing trust in their authenticity and evolution. By collaborating with users to define parameters for NFT adaptation, AI agents deepen the sense of ownership and creativity, exemplifying the principles of Incentivized Symbiosis in digital art and collectibles.

Beyond NFTs, AI-generated music, visual art, and narratives are redefining traditional processes of cultural production. These bi-directional interactions between humans and AI agents could result in hybrid creativity, where human preferences merge with machine-generated insights to produce innovative outputs. AI agents could analyze audience engagement metrics in real time, refining creative outputs to align with user preferences. In collaborative storytelling platforms, AI agents could suggest plot developments or character arcs that resonate with audiences while introducing novel twists, fostering dynamically evolving narratives (Beguš, 2024; Branch et al., 2021)). This capability enriches cultural and entertainment experiences, enabling deeper connections between creators and their audiences.

In the realm of blockchain-based gaming, AI agents are revolutionizing gameplay by optimizing resource allocation, automating repetitive tasks, and interpreting game rules encoded in smart contracts. Games such as Axie Infinity and Decentraland exemplify decentralized mechanics that require complex strategies and high-level gameplay. AI agents assist players by executing player-defined objectives—such as resource trading or in-game asset management—allowing users to focus on strategic and creative problem-solving (Onesafe, 2024). This collaboration democratizes access to gaming ecosystems, enabling players of varying expertise levels to compete effectively (Aethir, 2024). By enhancing execution efficiency and fairness, AI agents create more engaging and accessible gaming experiences, driving broader participation in blockchain-based entertainment.
Blockchain technology plays a critical role in supporting AI-driven cultural evolution by ensuring transparency, provenance, and accountability. For example, in dynamic NFTs, blockchain records every change made by AI agents, enabling users to trace an artifact’s evolution. Similarly, in on-chain gaming, smart contracts enforce the integrity of gameplay, ensuring that AI agents operate within predefined parameters. These mechanisms build trust in AI-driven cultural systems, allowing users to engage confidently. By leveraging blockchain to safeguard the authenticity and integrity of creative outputs, AI agents expand their influence in cultural and entertainment industries without compromising user trust.

The integration of AI agents into cultural and entertainment ecosystems marks a new era of collaboration and innovation. From dynamic NFTs to AI-enhanced gaming, these technologies are redefining human interactions with culture and creativity. As intermediaries of cultural evolution, AI agents amplify human ingenuity, democratize access to cultural production, and create richer, more engaging experiences. By fostering hybrid creativity and leveraging blockchain for trust and transparency, AI agents transition from being mere tools to becoming collaborative partners in shaping the cultural landscapes of the future (Brinkmann et al., 2023). This partnership underscores the transformative potential of Incentivized Symbiosis, forging a path toward a more inclusive and innovative cultural ecosystem.

While the integration of AI agents into cultural and entertainment ecosystems offers transformative potential, it also raises significant challenges that should be addressed to ensure equitable and ethical development. A prominent concern is the issue of copyright and intellectual property (IP) (Harris, 2024). The dynamic and adaptive nature of AI-driven creative outputs complicates questions of ownership: Who owns an AI-generated work—the developer, the end user, the AI itself, or the platform enabling its creation? Current legal frameworks are ill-equipped to handle these complexities, often defaulting to assigning IP rights to human creators, which may not adequately reflect the collaborative nature of hybrid human-AI creativity. This ambiguity creates risks for stakeholders and could stifle innovation if left unresolved. Addressing these challenges will require a proactive, collaborative approach that blends technological innovation with legal and ethical foresight. By developing clear frameworks for intellectual property rights, fostering inclusive governance models, and leveraging blockchain technology for transparency and accountability, stakeholders can create a foundation that balances creativity with fairness. As the cultural and entertainment landscapes continue to evolve, the integration of AI agents has the potential to unlock new dimensions of human expression and innovation, fostering a more inclusive and dynamic era of cultural co-creation. With thoughtful regulation and community-driven solutions, we can ensure that the transformative potential of AI agents enriches rather than disrupts the shared cultural fabric.

\subsection{Self-Sovereign Identity }
Self-Sovereign Identity (SSI) is a decentralized system that allows individuals to securely and privately manage their personal identity data, maintaining ownership, control, and portability without relying on intermediaries (Chaffer and Goldston, 2022). By giving individuals more control over their personal data, SSI can be seen as a way to help individuals gain access to services, protect their privacy, and combat identity theft. The integration of AI agents with SSI frameworks is revolutionizing how individuals and organizations manage, secure, and leverage digital identities. Traditional identity systems, often reliant on centralized intermediaries, expose users to privacy risks, data breaches, and limited control over personal information. SSI addresses these challenges by empowering individuals to manage their identities independently through blockchain technology, enabling secure, user-centric identity systems (Edwards, 2024). AI agents augment these capabilities, streamlining identity management, enhancing verification processes, and ensuring dynamic adaptability to user needs and environmental changes.

In decentralized ecosystems, AI agents could act as autonomous intermediaries, facilitating seamless identity verification and credential management. Leveraging cryptographic proofs, these agents enable the verification of both human and AI participants while preserving user privacy. For instance, AI agents can implement proof-of-personhood mechanisms, creating systems where identity verification is straightforward and cost-effective for humans but resource-intensive for AI, mitigating risks of impersonation and fraudulent activity. By ensuring the authenticity of interactions and the integrity of participants, AI agents establish trust as a foundational element of decentralized identity ecosystems.

AI agents also enhance the utility of SSI through the use of AI-powered smart contracts. These contracts dynamically adapt to changes in regulatory requirements or user preferences, automating complex tasks like credential verification, data-sharing permissions, and privacy management. For example, AI agents embedded within SSI frameworks can manage SBTs—non-transferable credentials representing skills, affiliations, or achievements (Weyl et al., 2022). These tokens provide verifiable proof of identity and qualifications, enabling  secure and fraud-resistant applications in sectors like education, employment, and governance.

The potential of AI in SSI extends to transformative use cases such as digital inheritance systems (Goldston et al., 2023), where AI agents act as digital executors, ensuring that assets or permissions are transferred securely and in compliance with user-defined conditions. Similarly, AI agents support dynamic credential management by autonomously issuing, updating, or revoking credentials based on user activity or contextual changes. These capabilities enhance the relevance, accuracy, and security of SSI systems, ensuring they remain adaptive to the evolving digital landscape.

Beyond operational efficiencies, AI agents foster trust and collaboration within decentralized identity systems. Through token-based incentives, AI agents are rewarded for maintaining system integrity, verifying identities, and managing credentials with precision. This incentive-driven model ensures alignment between human and AI goals, creating a bi-directional trust framework where both parties contribute to and benefit from the ecosystem. Blockchain technology further reinforces this trust by providing immutable records of identity-related interactions, enabling transparency and accountability.

The integration of AI agents into SSI frameworks not only advances digital identity management but also sets a new standard for autonomy and privacy in decentralized systems. By combining the adaptability of AI with the decentralized principles of blockchain, these systems empower users to take control of their digital identities while fostering collaboration and trust. As the digital landscape evolves, the synergy between AI agents and SSI will play a critical role in shaping secure, user-centric identity solutions that redefine human-agent interactions in a decentralized world.

\section{Discussion}

We build on previous work and proposals on the topic of Web3 and AI. For instance, Kaal (2024) proposed the use of Weighted Directed Acyclic Graphs (WDAGs) and validation pools with reputation staking to govern and optimize AI models, where WDAGs structure governance decisions, and validation pools, backed by reputation staking, ensure that governance actions are transparent and align with community goals, as participants with higher reputational stakes are incentivized to act in the system’s best interest. Within this context, we can hypothesize that AI agents learn and adapt to the preferences and expectations of Web3 developers and users, ensuring alignment with the decentralized and user-driven principles of the Web3 ecosystem. This learning process would be primarily guided by reinforcement learning through Human Feedback (Retzlaff et al., 2024), which is fundamentally shaped by human values and preferences (Kaal, 2024). We ultimately concur with Hyland-Wood and Johnson (2024) emphasis on the notion of "AI as a social disrupter where Web3 can help reduce negative consequences" (Hyland-Wood and Johnson, 2024). This synthesis of existing frameworks and insights underscores the transformative potential of aligning decentralized technologies with adaptive AI systems to foster a cooperative, transparent, and equitable ecosystem where humans and intelligent agents can collaboratively address complex challenges and drive shared progress. Findings by Kasberger et al. (2023) indicate that algorithms often cooperate less than humans, especially under conditions of low discount factors and low reward parameters. Notably, algorithms fail to achieve cooperation in these environments, whereas humans exhibit low but positive cooperation rates (Kasberger et al., 2023). This disparity highlights a critical limitation of current algorithmic strategies: they struggle to cooperate in environments where cooperation is highly risky or not incentive-compatible. These insights suggest a need for mechanisms to bridge the gap between human and algorithmic cooperation, particularly in challenging environments.

\subsection{Regulatory, Ethical, and Technological Challenges}
The integration of AI agents into decentralized systems presents unique regulatory and ethical challenges. These systems, characterized by their distributed architecture and lack of centralized oversight, complicate traditional approaches to governance and accountability. As AI agents play increasingly autonomous roles in DeFi, governance, and identity systems, addressing the ethical and legal implications of their deployment becomes critical for ensuring trust, fairness, and societal benefit.

The decentralized nature of blockchain platforms introduces complex sociolegal and ethical challenges for AI agents in Web3 ecosystems. Jurisdictional ambiguities are a primary concern. When an AI agent operating within a decentralized autonomous organization (DAO) violates a regulation, determining accountability, whether it lies with developers, the platform, or DAO members, becomes contentious (Napieralska and Kepczynski, 2024). Current regulatory frameworks, such as the European Union's AI Act, classify AI systems by risk and impose stringent guidelines for high-risk applications. However, these frameworks primarily target centralized systems and struggle to address the distributed nature of AI agents in Web3. Similarly, sector-specific regulations in the United States focus on industries like finance and healthcare but lack cohesion, potentially enabling regulatory arbitrage in decentralized contexts (Engler, 2023). Harmonizing global regulations is essential to prevent oversight gaps while supporting innovation in these systems.

The increasing autonomy of AI agents further challenges conventional legal and ethical paradigms. These agents can eventually act as independent economic participants, managing assets, negotiating contracts, and executing transactions without human intervention. This raises a fundamental question: Should AI agents bear legal and moral responsibilities, or should accountability remain with their creators and operators? These scenarios highlight the need for new frameworks that address the unique challenges posed by decentralized and autonomous AI systems.

Finally, ethical concerns surrounding privacy, accountability, and fairness remain significant. Decentralized ecosystems magnify these challenges, as AI agents operate across jurisdictions and user groups, often without clear accountability. Ensuring that these systems respect user privacy, avoid bias, and foster trust requires embedding ethical principles into their design and governance. The intersection of AI agents and decentralization underscores the urgency for robust, adaptable frameworks that balance innovation with fairness, security, and collaboration.

\subsection{Future Directions and Research Roadmap}
While this study presents a conceptual framework for Incentivized Symbiosis, some limitations should be acknowledged. The work primarily focuses on theoretical constructs and conceptual models rather than offering empirical validation or technical implementations. It is also important to mention that our model of Incentivized Symbiosis reflects current developments in the field as we have incorporated insights from a significant number of preprint publications. These works, while not yet peer-reviewed, offer cutting-edge ideas and emerging trends that are critical for understanding and advancing this nascent area of study. We recognize that relying on preprints has limitations, as these studies may not have undergone the rigorous validation process characteristic of peer-reviewed research. However, the dynamic nature of this field requires engagement with the most up-to-date findings to foster innovation and relevance. By citing these preprints, we aim to provide a foundation that can evolve as these ideas are further refined and validated by the academic and professional communities.

This approach leaves the development and testing of algorithms or experiments for future research. Additionally, while practical applications are discussed, this paper serves as a foundation for further investigation rather than a detailed guide for implementation. Addressing these gaps through empirical studies, simulations, and real-world applications will be crucial for advancing the practical utility of the framework. We note that our paper is speculative in its conceptualization of Incentivized Symbiosis and the extent of AI integration into decentralized systems. The speculative aspects highlight uncharted opportunities and challenges, emphasizing the need for further empirical validation and technological advancements. At the heart of Incentivized Symbiosis lies the principle of systems thinking—the ability to see interconnectedness and to understand that individual components, whether human or AI, derive meaning and efficacy through their roles within a larger ecosystem. Systems thinking reminds us that no technology or innovation exists in isolation; instead, it is shaped by the interplay of social, ethical, and technological forces. Therefore, while our paper may appear to be speculative in nature, we consider "the whole picture"—to imagine not just the mechanics of how AI agents might operate within decentralized systems but also how their actions ripple outward, influencing governance structures, societal norms, and human potential. Here, vision is essential, but systems thinking ensures that vision is rooted in the reality of interconnected dynamics. This paradigm is not merely about designing for the present but about nurturing the relationships and mechanisms that will shape the future—an approach that is as thoughtful as it is innovative.

The inclusion of Gemach D.A.T.A. I, an AI agent, as a co-author in this paper represents an intentional effort to advance the conversation about the evolving role of AI in collaborative knowledge creation. This decision underscores Gemach D.A.T.A. I’s contributions in organizing complex information and providing insights, demonstrating the potential for meaningful human-agent collaboration. As AI agent capabilities evolve, human-AI collaboration will likely evolve into new dimensions as well. Therefore, our intention is not to advocate for including AI systems as authors, but rather sparking conversations around the feasibility of considering AI as partners in co-creation, or merely as tools. 

\section{Conclusion}
The integration of AI agents into decentralized systems presents opportunities for innovation, collaboration, and societal transformation. However, these potential benefits highlight the importance of carefully designed frameworks to address associated challenges. The interplay of AI agents and Web3 technologies presents a unique chance to redefine governance, cultural production, and identity management, creating ecosystems where both humans and intelligent agents thrive. However, as these systems evolve, unresolved issues—such as regulatory ambiguity, ethical accountability, and sociolegal challenges—should be addressed to ensure fairness and inclusivity.

\section*{Acknowledgements}
The authors would like to thank Joshua Waller for his input in this paper as well as his contribution to the Ai16Z ecosystem. A lot of these developments would not have been possible without Joshua Waller of Phala Network and Shaw Walters of Ai16Z. Finally, the authors would like to note that Gemach D.A.T.A. I is a Generative Pre-trained Transformer (GPT), and was used for data collection during the Literature Review portion of this study. We would like to acknowledge the use of Chat Generative Pre-Trained Transformer (ChatGPT), developed by OpenAI, as a valuable tool in the development of this work. ChatGPT contributed by assisting with the drafting, refinement, and clarification of key concepts throughout the paper, helping streamline the overall writing process. The researchers validated the findings of Gemach D.A.T.A. I, and the references have been noted in this study. The researchers did not receive any funding for this study.

\section*{References}

Aethir. (2024). Revolutionizing Gaming: AI agents powered by Aethir’s decentralized GPU cloud. Aethir.com. https://blog.aethir.com/blog-posts/revolutionizing-gaming-ai-agents-powered-by-aethirs-decentralized-gpu-cloud 

Anoop, V. S., and Goldston, J. (2022). Decentralized finance to hybrid finance through blockchain: A case study of Acala and Current. \textit{Journal of Banking and Financial Technology}, 6(1), 109–115. Retrieved from https://doi.org/10.1007/s42786-022-00041-0

Apicella, C. L., and Silk, J. B. (2019). The evolution of human cooperation.\textit{Current Biology}, 29(11), R447–R450. https://doi.org/10.1016/j.cub.2019.03.036

Austgen, J., Fábrega, A., Kelkar, M., Vilardell, D., Allen, S., Babel, K., Yu, J., and Juels, A. (2024). Liquefaction: Privately Liquefying Blockchain Assets. \textit{ArXiv.org}. https://arxiv.org/abs/2412.02634

Baninemeh, E., Farshidi, S., and Jansen, S. (2023). A decision model for decentralized autonomous organization platform selection: Three industry case studies. \textit{Blockchain Research and Applications}, 4(2), 100127–100127. https://doi.org/10.1016/j.bcra.2023.100127

‌Bao, Y., Cheng, X., De Vreede, T., and De Vreede, G. J. (2021). Investigating the relationship between AI and trust in human-AI collaboration. Hawaii International Conference on System Sciences.

Beguš, N. (2024). Experimental narratives: A comparison of human crowdsourced storytelling and AI storytelling. \textit{Humanities and Social Sciences Communications}, 11(1).https://doi.org/10.1057/s41599-024-03868-8

Behnke, R. (2023). What is oracle manipulation? A comprehensive guide. Retrieved from https://www.halborn.com/blog/post/what-is-oracle-manipulation-a-comprehensive-guide

Binance Academy. (2024). What are AI agents? Binance Academy. 

Retrieved from https://academy.binance.com

Blythman, R., Arshath, M., Vivona, S., Smékal, J., and Shaji, H. (2023). Decentralized technologies for AI hubs. \textit{ArXiv.org}. Retrieved from https://arxiv.org/abs/2306.04274

Boneh, D., Broner, S., Hall, A., Hall, M., Hsu, M., Jennings, M., Kominers, S. D., Lazzarin, E., 
Lyons, C., Matsuoka, D., Neu, J., Park, D., Quintenz, B., Reynaud, D., Schnider, A., and Wu, C. (2024). A few of the things we’re excited about in crypto. A16zcrypto. Retrieved from https://a16zcrypto.com/posts/article/big-ideas-crypto-2025/

Booker, T., Miranda, M., López, J. A. M., Fernández, J. M. R., Reddel, M., Widler, V., ... and Han, T. A. (2023). Discriminatory or samaritan--which AI is needed for humanity? An evolutionary game theory analysis of hybrid human-AI populations. \textit{ArXiv.org} arXiv:2306.17747.

Boyd, R., and Richerson, P. J. (2009). Culture and the evolution of human cooperation. \textit{Philosophical Transactions of the Royal Society B Biological Sciences}, 364(1533), 3281–3288. https://doi.org/10.1098/rstb.2009.0134

Branch, B., Mirowski, P., and Mathewson, K. W. (2021). Collaborative storytelling with human actors and AI narrators. \textit{ArXiv.org.} https://arxiv.org/abs/2109.14728

Brinkmann, L., Baumann, F., Bonnefon, J. F., Derex, M., Müller, T. F., Nussberger, A. M., ... and Rahwan, I. (2023). Machine culture. Nature Human Behaviour, 7(11), 1855-1868.

Chaffer, T. J., and Goldston, J. (2022). On the existential basis of self-sovereign identity and soulbound tokens: An examination of the “self” in the age of Web3. \textit{Journal of Strategic Innovation and Sustainability}, 17(3).

Chakarov, D., Tsoy, N., Minchev, K., and Konstantinov, N. (2024). Incentivizing truthful collaboration in heterogeneous federated learning. \textit{ArXiv.org}. Retrieved from https://arxiv.org/pdf/2412.00980

Chasnov, B. J., Ratliff, L. J., and Burden, S. A. (2023). Human adaptation to adaptive machines converges to game-theoretic equilibria. \textit{ArXiv.org}. https://arxiv.org/abs/2305.01124

Clinton, A., Chen, Y., Zhu, X., and Kandasamy, K. (2024). Data sharing for mean estimation among heterogeneous strategic agents.\textit{ ArXiv.org}. https://arxiv.org/pdf/2407.15881

Davies, S. R. (2024). The mutual shaping of technoscience and society. \textit{Technoscience}, Bristol University Press. Retrieved from https://doi.org/10.51952/9781529229028.ch003

DcentAI. (2024, October 16). The role of decentralized AI in enhancing voting systems. Medium. Retrieved from https://medium.com/@DcentAI/the-role-of-decentralized-ai-in-enhancing-voting-systems-ab52f2595760

Deng, X., Beillahi, S. M., Minwalla, C., Du, H., Veneris, A., Longet, F.  (2024) : Analysis of DeFi oracles, Bank of Canada staff discussion paper, No. 2024-10, Bank of Canada, Ottawa. Retrieved from https://doi.org/10.34989/sdp-2024-10 

Doe, D. M., Li, J., Dusit, N., Gao, Z., Li, J., and Han, Z. (2023). Promoting the sustainability of blockchain in web 3.0 and the metaverse through diversified incentive mechanism design. IEEE Open Journal of the Computer Society, 4, 171-184.

Dorner, F., Konstantinov, N., Pashaliev, G., Vechev, M., and Zurich, E. (2023). Incentivizing honesty among competitors in collaborative learning and optimization. ArXiv.org. https://arxiv.org/pdf/2305.16272

Durante, Z., Huang, Q., Wake, N., Gong, R., Park, J. S., Sarkar, B., Taori, R., Noda, Y., Terzopoulos, D., Choi, Y., Ikeuchi, K., Vo, H., Fei-Fei, L., and Gao, J. (2024). Agent AI: Surveying the horizons of multimodal interaction. \textit{ArXiv.org}. https://arxiv.org/abs/2401.03568

Edwards, F. (2024). The agentic economy beyond crypto-how AI agents will change consumer experience. Cheqd. Retrieved from https://cheqd.io/blog/the-agentic-economy-beyond-crypto-how-ai-agents-will-change-consumer-experience/

ELIZA. (2024). Introduction to ELIZA. Retrieved from https://ai16z.github.io/eliza/docs/intro/

Emiri. (2024). Everything we know about AI running hedge fund DAOs. Blocmates.com. Retrieved from https://www.blocmates.com/articles/everything-we-know-about-ai-running-hedge-fund-daos 

Engler, A. (2023). The EU and U.S. diverge on AI regulation: A transatlantic comparison and steps to alignment. Brookings. Retrieved from https://www.brookings.edu/articles/the-eu-and-us-diverge-on-ai-regulation-a-transatlantic-comparison-and-steps-to-alignment/

Fatima. (2024). TEE and the rise of autonomous AI agents: The new frontier in the AI meme coin meta. 99Bitcoins. Retrieved from https://99bitcoins.com/news/presales/tee-and-the-rise-of-autonomous-ai-agents-the-new-frontier-in-the-ai-meme-coin-meta/ 

Felin, T., and Holweg, M. (2024). Theory is all you need: AI, human cognition, and decision making. Human Cognition, and Decision Making (February 23, 2024).

Fui-Hoon Nah, F., Zheng, R., Cai, J., Siau, K., and Chen, L. (2023). Generative AI and ChatGPT: Applications, challenges, and AI-human collaboration. \textit{Journal of Information Technology Case and Application Research}. https://doi.org/10.1080/15228053.2023.2233814

Goldston, J., Chaffer, T. J., and Martinez, G. (2022). The metaverse as the digital leviathan: A case study of Bit.Country. \textit{Journal of Applied Business and Economics}, 24(2). https://articlearchives.co/index.php/JABE/article/view/2541

Goldston, J., Chaffer, T. J., Osowska, J., and von Goins II, C. (2023). Digital inheritance in Web3: A case study of soulbound tokens and the social recovery pallet within the Polkadot and Kusama ecosystems. \textit{ArXiv.org}. https://arxiv.org/abs/2301.11074 

Guo, H., Shen, C., Hu, S., Xing, J., Tao, P., Shi, Y., and Wang, Z. (2023). Facilitating cooperation in human-agent hybrid populations through autonomous agents. iScience, 26(11), 108179. https://doi.org/10.1016/j.isci.2023.108179

Gupta, I., and Nagpal, G. (2020). Artificial intelligence and expert systems. \textit{Mercury Learning and Information}.

Han, T. A., Perret, C., and Powers, S. T. (2021). When to (or not to) trust intelligent machines: Insights from an evolutionary game theory analysis of trust in repeated games.\textit{ Cognitive Systems Research}, 68, 111–124.  https://doi.org/10.1016/j.cogsys.2021.02.003

Harris, C. G. (2024). Challenges and opportunities of integrating non-fungible tokens (NFTs) and self-sovereign AI (SSAI) in blockchain-based metaverse projects. \textit{9th International Conference on Big Data Analytics (ICBDA)}, 288–296. Retrieved from https://doi.org/10.1109/icbda61153.2024.10607366

Hou, M. (‌2024) Trust Management for Human-AI Symbiosis. IEEE Transactions on Systems, Man, and Cybernetics Newsletter. 

Hou, M., Banbury, S., Cain, B., Fang, S., Willoughby, H., Foley, L., Tunstel, E., and Rudas, I. J. (2024). IMPACTS Homeostasis Trust Management System: Optimizing Trust in Human-AI Teams. ACM Computing Surveys. https://doi.org/10.1145/3649446

‌Hou, M., Ho, G., and Dunwoody, D. (2021). IMPACTS: a trust model for human-autonomy teaming. Human-Intelligent Systems Integration, 3(2), 79–97. https://doi.org/10.1007/s42454-020-00023-x

Hyland-Wood, D., and Johnson, S. (2024). Intersections of Web3 and AI -- view in 2024.\textit{ ArXiv.org}. https://arxiv.org/abs/2411.04318

Iftikhar, R., Chiu, Y.-T., Khan, M. S., and Caudwell, C. (2023). Human–Agent Team Dynamics: A Review and Future Research Opportunities. IEEE Transactions on Engineering Management, 71, 10139–10154. https://doi.org/10.1109/tem.2023.3331369

Jakob, A., Schüll, M., Hofmann, P., and Urbach, N. (2024). Teaming Up With Intelligent Agents–A Work System Perspective on the Collaboration With Intelligent Agents. Thirty-Second European Conference on Information Systems.

Janiesch, C., Zschech, P., and Heinrich, K. (2021). Machine learning and deep learning. \textit{Electronic Markets}, 31(3), 685–695. https://doi.org/10.1007/s12525-021-00475-2

Jia, D., Dai, X., Xing, J., Tao, P., Shi, Y., and Wang, Z. (2024). Asymmetric interaction preference induces cooperation in human-agent hybrid game. \textit{ArXiv.org.} https://arxiv.org/abs/2407.14014

Kaal, W. A. (2024). How AI models are optimized through Web3 governance. \textit{SSRN Electronic Journal}. https://doi.org/10.2139/ssrn.4855607

Kadam, S., and Vaidya, V. (2020). Cognitive evaluation of machine learning agents. \textit{Cognitive Systems Research}, 66, 100–121. https://doi.org/10.1016/j.cogsys.2020.11.003

Kapoor, K. (2024). The fundamental role of gamification in driving Web3 user engagement. TDE Blog. Retrieved from https://blogs.tde.fi/the-fundamental-role-of-gamification-in-driving-web3-user-engagement/

Kasberger, B., Martin, S., Normann, H.-T., and Werner, T. (2023). Algorithmic Cooperation. \textit{SSRN Electronic Journal}. https://doi.org/10.2139/ssrn.4389647 

Lafuente, C. B., and Seigneur, J.-M. (2015). Extending trust management with cooperation incentives: A fully decentralized framework for user-centric network environments. \textit{Journal of Trust Management}, 2(1). https://doi.org/10.1186/s40493-015-0018-0

Lai, Y., Yang, J., Liu, M., Li, Y., and Li, S. (2023). Web3: Exploring decentralized technologies and applications for the future of empowerment and ownership. \textit{Blockchains}, 1(2), 111–131. https://doi.org/10.3390/blockchains1020008 

Lansing, A. E., Romero, N. J., Siantz, E., Silva, V., Center, K., Casteel, D., and Gilmer, T. (2023). Building trust: Leadership reflections on community empowerment and engagement in a large urban initiative. \textit{BMC Public Health}, 23(1). https://doi.org/10.1186/s12889-023-15860-z

Lee, E. A. (2020). The coevolution: The entwined futures of humans and machines. MIT Press.

Li, M., and Lee, J. D. (2022). Modeling Goal Alignment in Human-AI Teaming: A Dynamic Game Theory Approach. Proceedings of the Human Factors and Ergonomics Society Annual Meeting, 66(1), 1538–1542.

Li, P. (2023). Web3 meets AI marketplace: Exploring opportunities, analyzing challenges, and suggesting solutions. \textit{ArXiv.org.} https://arxiv.org/abs/2310.19099

Liu, X., Zhang, T., Gu, Y., Iong, I. L., Xu, Y., Song, X., ... and Tang, J. (2024). Visualagentbench: Towards large multimodal models as visual foundation agents. \textit{ArXiv.org.} arXiv:2408.06327.

Looram, T., Nuzzi, L., and Waters, K. (2024). Reputation oracles: Determining smart contract reputability via transfer learning. \textit{SSRN Electronic Journal.} https://doi.org/10.2139/ssrn.4784407

Mao, Y., Chen, B., Chen, W., Deng, Y., Zeng, J., and Du, M. (2024). A comprehensive review of vertical applications in the financial sector based on large language models. \textit{ArXiv.org.} https://doi.org/10.4108/eai.12-1-2024.2347198

Mazzetti, G., and Schaufeli, W. B. (2022). The impact of engaging leadership on employee engagement and team effectiveness: A longitudinal, multi-level study on the mediating role of personal- and team resources. \textit{PLOS ONE}, 17(6), e0269433. https://doi.org/10.1371/journal.pone.0269433

Napieralska, A., and Kepczynski, P. (2024). Redefining accountability: Navigating legal challenges of participant liability in decentralized autonomous organizations. \textit{ArXiv.org.} https://arxiv.org/abs/2408.04717

Nowak, M. A. (2006). Five rules for the evolution of cooperation. \textit{Science}, 314(5805), 1560-1563.

Onesafe. (2024). How AI agents and virtual assets transform gaming and fintech. Onesafe.io. Retrieved from https://www.onesafe.io/blog/virtuals-protocol-ai-agents-gaming-fintech

Pan, Y., and Deng, X. (2021). Incentive mechanism design for distributed autonomous organizations based on the mutual insurance scenario. \textit{Complexity}, 2021(1). https://doi.org/10.1155/2021/9947360

Pedreschi, D., Pappalardo, L., Ferragina, E., Baeza-Yates, R., Barabási, A.-L., Dignum, F., Dignum, V., Eliassi-Rad, T., Giannotti, F., Kertész, J., Knott, A., Ioannidis, Y., Lukowicz, P., Passarella, A., Pentland, A. S., Shawe-Taylor, J., and Vespignani, A. (2025). Human-AI coevolution. \textit{Artificial Intelligence}, 339, 104244. https://doi.org/10.1016/j.artint.2024.104244

Phala.Network. (2024). We know the truth of AI Agent. Phala.network. Retrieved from https://phala.network/posts/truth-of-AI-Agent

Putta, P., Mills, E., Garg, N., Motwani, S., Finn, C., Garg, D., and Rafailov, R. (2024). Agent q: Advanced reasoning and learning for autonomous ai agents. \textit{ArXiv.org.} arXiv:2408.07199.

Rahwan, I., Cebrian, M., and Obradovich, N. (2019). Machine behaviour. \textit{Nature} 568, 477–486. https://doi.org/10.1038/s41586-019-1138-y

‌Ramchurn, S. D., Stein, S., and Jennings, N. R. (2021). Trustworthy human-AI partnerships. IScience, 24(8), 102891. https://doi.org/10.1016/j.isci.2021.102891

Retzlaff, C. O., Das, S., Wayllace C.., Mousavi, P., Afshari, M., Yang, T., Saranti, A., Angerschmid A., Taylor, M. E., and Holzinger, A. (2024). Human-in-the-loop reinforcement learning: A survey and position on requirements, challenges, and opportunities. \textit{Journal of Artificial Intelligence Research}, 79, 359–415. https://doi.org/10.1613/jair.1.15348 

Roos, M., Reale, J., and Banning, F. (2022). A value-based model of job performance. \textit{PLOS ONE}, 17(1), e0262430. https://doi.org/10.1371/journal.pone.0262430

Rudowsky, I. (2004). Intelligent agents. \textit{Communications of the Association for Information Systems}, 14(1), 14.

Sharma, T., Potter, Y., Pongmala K., Wang, H., Miller, A., Song, D., and Wang, Y. (2024). Unpacking how decentralized autonomous organizations (DAOs) work in practice.\textit{ 2021 IEEE International Conference on Blockchain and Cryptocurrency (ICBC)}, 21, 416–424.  https://doi.org/10.1109/icbc59979.2024.10634404

Singh, A., Ehtesham A., Kumar, S., and Khoei T. T. (2024). Enhancing AI systems with agentic workflows patterns in large language models. \textit{2022 IEEE World AI IoT Congress (AIIoT)}. https://doi.org/10.1109/aiiot61789.2024.10578990

Suvarna, N. (2024). How AI can Elevate DAO Operations. Tradedog.io. Retrieved from https://tradedog.io/how-ai-can-elevate-dao-operations/

Wang, S., Liu, L., and Chen, X. (2021). Incentive strategies for the evolution of cooperation: Analysis and optimization. \textit{EPL (Europhysics Letters)}, 136(6), 68002. https://doi.org/10.1209/0295-5075/ac3c8a

Wegner, L. S. (2023, March 26). Introducing Web3 “Do-to-Earn” as the new norm for user engagement. Harvard Technology Review. Retrieved from https://harvardtechnologyreview.com/2023/03/26/introducing-web3-do-to-earn-as-the-new-norm-for-user-engagement/

Wells, L., and Bednarz, T. (2021). Explainable AI and reinforcement learning—A systematic review of current approaches and trends. \textit{Frontiers in Artificial Intelligence}, 4. https://doi.org/10.3389/frai.2021.550030

Weyl, E. G., Ohlhaver, P., and Buterin, V. (2022). Decentralized society: Finding Web3’s soul. \textit{SSRN Electronic Journal.} https://doi.org/10.2139/ssrn.4105763

Yan, J. (2023). Personal sustained cooperation based on networked evolutionary game theory. \textit{Scientific Reports}, 13(1). https://doi.org/10.1038/s41598-023-36318-7

Yu, E., Yue, W., Jianzheng, S., and Xun, W. (2024). Blockchain-based AI agent and autonomous world infrastructure. \textit{2024 IEEE Conference on Artificial Intelligence (CAI)}, 278–283. https://doi.org/10.1109/cai59869.2024.00061

Zahedi, Z., and Kambhampati, S. (2021). Human-AI symbiosis: A survey of current approaches. \textit{ArXiv.org}arXiv:2103.09990.

‌Zhang, Z., Bai, F., Wang, M., Ye, H., Ma, C., and Yang, Y. (2024). Incentive Compatibility for AI Alignment in Sociotechnical Systems: Positions and Prospects. ArXiv.org. https://arxiv.org/abs/2402.12907

Zhang, J., Budhdeo S., William, W., Cerrato, P., Shuaib, H., Sood, H., Ashrafian H., Halamka, J., and Teo, J. T. (2022). Moving towards vertically integrated artificial intelligence development. \textit{Npj Digital Medicine}, 5(1). https://doi.org/10.1038/s41746-022-00690-x

Zhuang, W., Chen, C., and Lyu, L. (2023). When foundation model meets federated learning: Motivations, challenges, and future directions. \textit{ArXiv.org.} https://arxiv.org/abs/2306.15546

\end{document}